\documentclass[aps,prl,amsmath,amssymb,twocolumn,showpacs,superscriptaddress]{revtex4-1}

\usepackage{dcolumn}
\usepackage{bm}
\usepackage{graphicx}
\usepackage{epstopdf}
\usepackage{wrapfig}
\usepackage{hyperref}
\usepackage{array} 
\usepackage{listings}
\usepackage[para,online,flushleft]{threeparttablex}

\usepackage{tikz}
\usetikzlibrary{arrows}
\usetikzlibrary{shadings}

\begin{document}

\preprint{COLLAPS/Ca spin and moments}

\title{Ground-State Electromagnetic Moments of Calcium Isotopes}

\author{R. F. Garcia Ruiz}\email{ronald.fernando.garcia.ruiz@cern.ch}\affiliation{KU Leuven, Instituut voor Kern-en Stralingsfysica, B-3001 Leuven, Belgium}
\author{M. L. Bissell}\affiliation{KU Leuven, Instituut voor Kern-en Stralingsfysica, B-3001 Leuven, Belgium}
\author{K. Blaum}\affiliation{Max-Plank-Institut f\"ur Kernphysik, D-69117 Heidelberg, Germany}
\author{N. Fr$\ddot{\text{o}}$mmgen}\affiliation{Institut f\"ur Kernchemie, Universit\"at Mainz, D-55128 Mainz, Germany}
\author{M. Hammen}\affiliation{Institut f\"ur Kernchemie, Universit\"at Mainz, D-55128 Mainz, Germany}
\author{J.~D.~Holt}\affiliation{Institut f\"ur Kernphysik, Technische Universit\"at Darmstadt, 64289 Darmstadt, Germany}\affiliation{Extreme Matter Institute EMMI, GSI Helmholtzzentrum f\"ur Schwerionenforschung GmbH, 64291 Darmstadt, Germany}\affiliation{TRIUMF, 4004 Wesbrook Mall, Vancouver, British Columbia, V6T 2A3, Canada}
\author{M. Kowalska}\affiliation{CERN, European Organization for Nuclear Research, Physics Department, CH-1211 Geneva 23, Switzerland}
\author{K. Kreim}\affiliation{Max-Plank-Institut f\"ur Kernphysik, D-69117 Heidelberg, Germany}
\author{J. Men\'endez}\affiliation{Institut f\"ur Kernphysik, Technische Universit\"at Darmstadt, 64289 Darmstadt, Germany}\affiliation{Extreme Matter Institute EMMI, GSI Helmholtzzentrum f\"ur Schwerionenforschung GmbH, 64291 Darmstadt, Germany}\affiliation{Department of Physics, University of Tokyo, Hongo, Tokyo 113-0033, Japan}
\author{R. Neugart}\affiliation{Max-Plank-Institut f\"ur Kernphysik, D-69117 Heidelberg, Germany}\affiliation{Institut f\"ur Kernchemie, Universit\"at Mainz, D-55128 Mainz, Germany}
\author{G. Neyens}\affiliation{KU Leuven, Instituut voor Kern-en Stralingsfysica, B-3001 Leuven, Belgium}
\author{W. N$\ddot{\text{o}}$rtersh$\ddot{\text{a}}$user}\affiliation{Institut f\"ur Kernphysik, Technische Universit\"at Darmstadt, 64289 Darmstadt, Germany}
\author{F. Nowacki}\affiliation{IPHC, IN2P3-CNRS and Universite Louis Pasteur, F-67037 Strasbourg, France}
\author{J. Papuga}\affiliation{KU Leuven, Instituut voor Kern-en Stralingsfysica, B-3001 Leuven, Belgium}
\author{A. Poves}\affiliation{Departamento de F\'isica Te\'orica and IFT-UAM/CSIC, Universidad Aut\'onoma de Madrid, E-28049 Madrid, Spain}
\author{A. Schwenk}\affiliation{Institut f\"ur Kernphysik, Technische Universit\"at Darmstadt, 64289 Darmstadt, Germany}\affiliation{Extreme Matter Institute EMMI, GSI Helmholtzzentrum f\"ur Schwerionenforschung GmbH, 64291 Darmstadt, Germany}
\author{J. Simonis}\affiliation{Institut f\"ur Kernphysik, Technische Universit\"at Darmstadt, 64289 Darmstadt, Germany}
\affiliation{Extreme Matter Institute EMMI, GSI Helmholtzzentrum f\"ur Schwerionenforschung GmbH, 64291 Darmstadt, Germany}
\author{D. T. Yordanov}\affiliation{Max-Plank-Institut f\"ur Kernphysik, D-69117 Heidelberg, Germany}

\begin{abstract}

High-resolution bunched-beam collinear laser spectroscopy was used to measure the optical hyperfine spectra of the $^{43-51}$Ca isotopes. The ground state magnetic moments of $^{49,51}$Ca and quadrupole moments of $^{47,49,51}$Ca were measured for the first time, and the $^{51}$Ca ground state spin $I=3/2$ was determined in a model-independent way. Our results provide a critical test of modern nuclear theories based on shell-model calculations using phenomenological as well as microscopic interactions. The results for the neutron-rich isotopes are in excellent agreement with predictions using interactions derived from chiral effective field theory including three-nucleon forces, while lighter isotopes illustrate the presence of particle-hole excitations of the $^{40}$Ca core in their ground state.

\end{abstract}

\pacs{Collinear laser spectroscopy, exotic nuclei, moments and spin, calcium isotopes}

\maketitle

The existence of doubly magic nuclei has played a key role in our understanding of nuclear structure. They have been the basis to develop the shell model, and are an ideal probe to test our knowledge of nuclear interactions by comparing experimental data with shell-model predictions \cite{brown01,caurier05}. Such shell-model calculations depend on the effective Hamiltonian used, a suitable valence space to capture the low-energy degrees of freedom, and consistent effective operators. Although effective charges and $g$-factors are widely used in shell-model calculations, they are not completely understood. Furthermore, their orbital \cite{valiente09} and valence-space  \cite{stetcu13} dependence and connection to two-body currents (meson-exchange currents), known to be important for magnetic moments in light nuclei \cite{pastore13}, is under discussion.

Having a closed proton shell, $Z=20$, and two naturally occurring
doubly magic isotopes, $^{40}$Ca and $^{48}$Ca, the calcium isotopic
chain has always been considered a prime benchmark for nuclear
structure, both from a theoretical \cite{talmi62} and an experimental
perspective \cite{kingbook}. 
Recently, special attention has turned to the evolution of
the structure beyond $N=28$, where additional shell closures have been suggested at $N=32$ \cite{wienholtz13} and $N=34$ \cite{nature54Ca}. These neutron-rich calcium isotopes have also gained exceptional interest
from the theoretical side \cite{holt12,hagen12,roth12,soma13,holt14}, as their properties reveal new aspects of nuclear forces, in particular regarding the role of three-nucleon (3N) forces \cite{holt12,hagen12} (for a review on 3N forces see \cite{hammer13}).

Spectroscopic properties in the Ca region are well described by phenomenological shell-model interactions, such as KB3G \cite{kb3} and GXPF1A \cite{gxpf}. These interactions start from a $^{40}$Ca core and two-nucleon (NN) forces and refit part of the
interactions to experimental data in the $pf$ shell, to compensate for
neglected many-body effects (both due to 3N and many-body correlations) \cite{caurier05}. Normal and super-deformed bands in $^{40}$Ca have been understood as due to particle-hole excitations of protons and neutrons from the $sd$-shell into the $pf$-orbits, and have been well-described using the SDPF.SM shell model interaction starting from a virtual $^{28}$Si core \cite{caurier07}. These calculations showed that the $^{40}$Ca ground state is very correlated. In the last years,
valence-shell interactions have been derived from NN and 3N
forces based on chiral effective field
theory \cite{holt12}, fitted only to few-nucleon
systems. Investigating the reliability of these microscopic NN+3N interactions is a matter of general interest as they have direct implications for the modeling of astrophysical systems \cite{helbert13}. The NN+3N interactions provide a good description of the shell structure and the spectra of neutron rich calcium isotopes in an extended valence space ($fpg_{9/2}$) \cite{holt14}.  
The electric quadrupole ($E2$) transitions obtained from both phenomenological and microscopic interactions exhibit good agreement using the neutron effective charge: $e_{n}=0.5 e$. On the other hand, phenomenological interactions and NN+3N disagree in effective nucleon $g$-factors needed to reproduce the magnetic ($M1$) transition strengths \cite{holt14}.

Despite the remarkable differences, both phenomenological and microscopic NN+3N interactions give a similar description of neutron separation (binding) energies and low-lying excitation energies of Ca isotopes from $N=22$ up to $N=32$ \cite{wienholtz13,nature54Ca,gallant}. As illustrated in \cite{isacker11}, such observables might however be insensitive to cross-shell correlations. Therefore, there is a need to measure additional observables like electromagnetic moments, which further test the above models and might provide a deeper insight to developing improved shell-model interactions. 

Magnetic moments and  $g$-factors, $g=\mu/(I \mu_N)$, of isotopes near shell closures are very sensitive to the occupancy of particular orbitals by valence
particles (or holes). The quadrupole moments on the other hand are
directly sensitive to nuclear shell structure \cite{townes49}. While
the terms \textit{closed shell} or \textit{magic number} may lack a
rigid definition, the electromagnetic moments provide a more direct
probe of the structure involved including cross-shell effects
\cite{stoyle56,neyens03}.

In this Rapid Communication, we report the first measurements of the quadrupole
moment of the closed shell $-1$ isotope $^{47}$Ca, and the quadrupole
and magnetic moment of the closed-shell +1 isotope
$^{49}$Ca \footnote{A previous value for the magnetic moment of
$^{49}$Ca was suggested from a partial measurement of its hfs and
assuming a value for its unknown isotope
shift \cite{vermeeren93}.}. Also the magnetic and quadrupole moments of
$^{51}$Ca, having a single-hole with respect to the new $N=32$
subshell closure, are presented, as well as its ground state (g.s.) spin. The experimental data are compared to shell-model calculations using phenomenological interactions, and to calculations including 3N forces based on chiral effective field theory.

At ISOLDE, CERN, exotic Ca isotopes were produced from nuclear
reactions induced by a high-energy proton beam (1.4 GeV; pulses of 2 $\mu$C typically every 2.4 s) impinging on a uranium carbide target. High selectivity for
the Ca reaction products was accomplished by laser
ionization \cite{fedosseev}. Ions were extracted from the ion
source and accelerated up to 30 keV or 40 keV to be mass separated, after which
they were injected into the ISOLDE radio-frequency quadrupole (RFQ) beam cooler, ISCOOL \cite{mane09}. Ions
were trapped for approximately 50 ms, and extracted bunches of 5 $\mu$s temporal
width were distributed to a dedicated beam line for collinear laser
spectroscopy experiments (COLLAPS). At COLLAPS, the ion beam was
superimposed with a continuous wave (CW) laser beam from a frequency-doubled Ti:Sa
laser, providing a 393-nm laser wavelength to excite the
$4s$ $^2S_{1/2} \rightarrow 4p$ $^2P_{3/2}$ transition in Ca$^{+}$. The
laser frequency was locked to a Fabry-Perrot interferometer, which was
in turn locked to a polarization-stabilized HeNe laser, reducing the
laser frequency drift to $< 10$~MHz per day.

\begin{figure}[t]
\includegraphics[scale=0.3]{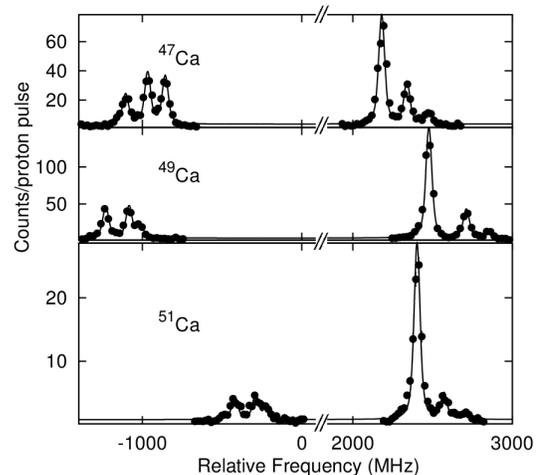}
\vspace{-0.44cm}
\caption{Examples of hfs spectra measured for the Ca isotopes in the 
393 nm $4s^2S_{1/2} \rightarrow 4p^2P_{3/2}$ ionic transition. The
lines show the fit with a Voigt profile. Frequency values are relative
to the centroid of $^{43}$Ca.\label{spectra}}
\end{figure}

By changing the ion velocity, and thereby Doppler tuning the laser
frequency in the ion rest frame, hyperfine structure (hfs) components
could be scanned. Fluorescence photons were detected by a set of four
photomultipliers (PMT) at the end of the beam line (see
Refs.~\cite{papuga13,kim13} for details). By only accepting signals
from the PMT whilst the ion bunch passed in front of them,
background from scattered laser light and PMT dark counts was reduced
by a factor of $\sim$ 10$^{4}$. Sample hfs spectra measured during the
experiment are shown in Fig.~\ref{spectra}. The magnetic hfs
constants, $A(^2S_{1/2})$, $A(^2P_{3/2})$, and quadrupole hfs
constants, $B(^2P_{3/2})$, were extracted from the fit of Voigt
profiles to the experimental spectra by using a $\chi^2$-minimization
technique as explained, e.g., in Ref. \cite{kowalska}. The values are listed
in Table \ref{HFStable}.  Only $^{43}$Ca has been studied before in this ionic transition, and our values are in agreement  within 1.1 standard deviations. For $^{41}$Ca and $^{45}$Ca, two measurements of the quadrupole hfs constants have been
reported in the atomic level system, both relative to that of
$^{43}$Ca \cite{bergman80,arnold83}, yielding the ratios
$B(^{41}\text{Ca})/B(^{43}\text{Ca})=1.63(1)$, and
$B(^{45}\text{Ca})/B(^{43}\text{Ca})=-0.94(27)$. The $B$-factor ratio equals
the ratio of the quadrupole moments. Thus we can compare the ratio of our $B$-values, measured in the ionic system $B(^{45}\text{Ca})/B(^{43}\text{Ca})=-0.74(31)$, to the latter value. They are in agreement within the error bars.

\begin{table}[t]
\begin{threeparttable}
\caption{Hyperfine structure values obtained from the fit to the 
experimental data compared to previous measurements.\label{HFStable}}
\begin{ruledtabular}
\begin{tabular}{c|llldc}
$A$ & $I^{\pi}$ & \multicolumn{1}{c}{$A(^2S_{1/2})$} & \multicolumn{1}{c}{$A(^2P_{3/2})$} & \multicolumn{1}{c}{$B(^2P_{3/2})$} & \multicolumn{1}{c}{Ref.} \\
&& \multicolumn{1}{c}{(MHz)} & \multicolumn{1}{c}{(MHz)} & \multicolumn{1}{c}{(MHz)} & \\\hline
43 & 7/2$^{-}$ & -806.87(42) & -31.10(30) & -4.2 (1.3) & \\
& & -806.40207160(8) & & & \cite{arbes}\\
& & -805(2) & -31.9(2) & -6.7(1.4) & \cite{silveransZD}\\
& & & -31.0(2) & -6.9(1.7) & \cite{norte98}\\ 
\hline
45 & 7/2$^{-}$ & -811.99(44) & -31.43(19) & 3.1(1.0) & \\
47 & 7/2$^{-}$ & -860.96(28) & -33.33(13) & 12.68(96) & \\
49 & 3/2$^{-}$ & -1971.02(30) & -75.98(11) & -5.53(40) & \\
51 & 3/2$^{-}$ & -1499.22(94) & -58.15(54) & 5.4(1.8) &	\\
\end{tabular}
\end{ruledtabular}
\end{threeparttable}
\end{table}

\begin{figure}[]
\scalebox{0.7}{\input{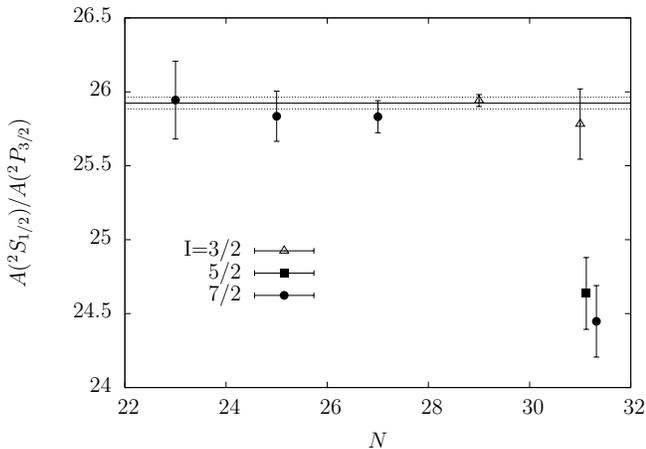}}
\vspace{-0.6cm}
\caption{Ratio between the hfs constants $A(^2P_{1/2})$ and $A(^2P_{3/2})$.
The continuous line shows the average value
$A(^2S_{1/2})/A(^2P_{3/2})=25.92(3)$. Hyperfine structure spectra of
$^{51}$Ca were fitted assuming different g.s. spin 
values of $I=3/2,5/2,7/2$.\label{spindetermine2}}
\end{figure}

The nuclear spin, $I$, is required to calculate each peak position in
the minimization procedure. Since a different set of hfs constants is
found for a given spin, the ratio, $R=A(^2P_{3/2})/A(^2S_{1/2})$, can
be used to determine the correct spin for each isotope, as this ratio
should be a constant over the entire isotopic chain (neglecting a possible small hyperfine anomaly). As it can be seen from
Fig.~\ref{spindetermine2}, the ratio $R$ remains constant along the Ca
isotopes up to $^{49}$Ca, using the earlier determined g.s. spins in the
fitting procedure. For $^{51}$Ca we assumed three possible spins for
its ground state and only when $I=3/2$ is used, the ratio of the
fitted hfs parameters is consistent with those from the other
isotopes. Thus $I=3/2$ is the g.s. spin of $^{51}$Ca, confirming
earlier tentative assignments \cite{perrot06,fornal08}, and in
agreement with expectations from the shell model.

Magnetic moments were extracted from the lower state magnetic hfs
constants, $A={\mu_I B_0/(IJ)}$, where $B_0$ is the magnetic field
produced by the electrons at the nucleus, and $J$ is the electronic total angular momentum. Since high-precision values
of $A(^2S_{1/2})=-806.40207160(8)$ MHz \cite{arbes} and
$\mu=-1.3173(6)$ \cite{olschewski72} are known for $^{43}$Ca, this isotope was used as a reference to calculate the other
magnetic moments from the measured $A$-values. The results are shown
in Table~\ref{QMtable}, where we compare our data to earlier reported
values for $^{45}$Ca and $^{47}$Ca.

\begin{table}[t]
\begin{threeparttable}
\caption{Quadrupole and magnetic moments obtained from the measured 
hfs constants (Table~\ref{HFStable}). The magnetic moments were
obtained using the reference isotope $^{43}$Ca, with
$A(^2P_{3/2})=-806.40207160(8)$ MHz \cite{arbes}. Quadrupole moments were
extracted using the calculated electric field gradient,
$eV_{JJ}=151.3(7)$ MHz/b \cite{sahoo09}. Data are compared to calculations using the NN+3N interaction.\label{QMtable}}
\begin{ruledtabular}
\begin{tabular}{llcllc}
$A$ & \multicolumn{1}{c}{$\mu(\mu_N)$} & \multicolumn{1}{c}{$\mu(\mu_N)$} & \multicolumn{1}{c}{$Q$ (b)} & \multicolumn{1}{c}{$Q$ (b)} & {Ref.} \\ 
&& \multicolumn{1}{c}{(NN+3N)} && \multicolumn{1}{c}{(NN+3N)} & \\\hline
41 & $-$1.594781(9) & & & & \cite{brun62} \\
 &  & & $-$0.080(8) & & \cite{arnold83} \\
43 &  & $-$1.56 & $-$0.028(9) & $-$0.0246 & This work \\
& $-$1.3173(6)\tnote{b}  & & &  & \cite{olschewski72} \\
& & & $-$0.043(9) & & \cite{silveransZD} \\
& & & $-$0.049(5) & & \cite{arnold83} \\
& & & $-$0.0408(8) & & \cite{sundholm} \\
& & & $-$0.0444(6) & & \cite{sahoo09} \\
45 & $-$1.3264(13) & $-$1.45 & +0.020(7) & +0.0252 & This work \\
& $-$1.3278(9) & &  & & \cite{arnold81} \\
 &  & & +0.046(14) & & \cite{arnold83} \\
47 & $-$1.4064(11) & $-$1.38 & +0.084(6) & +0.0856 & This work	\\
& $-$1.380(24) & & & & \cite{andl82} \\
49 & $-$1.3799(8) & $-$1.40 & $-$0.036(3) & $-$0.0422 & This work \\
51 & $-$1.0496(11) & $-$1.04 & $+0.036(12)$ & +0.0425 & This work \\
\end{tabular}
\end{ruledtabular}
\begin{tablenotes}
\raggedright
{\footnotesize 
\item[b] Reference value.
}
\end{tablenotes}
\end{threeparttable}
\end{table}

Quadrupole moments, $Q$, were obtained from the quadrupole hfs constant,
$B=e Q V_{JJ}$, with $e$ the electron charge, and $V_{JJ}$ the
electric field gradient (EFG) produced by the electrons at the
nucleus, the latter being isotope independent. To extract quadrupole moments from the measured
hfs $B$-parameters, a calculated value for the EFG, $e V_{JJ}=151.3(7)$
MHz/b, was taken from atomic-physics calculations based on
relativistic coupled-cluster theory (RCC) \cite{sahoo09}. Independent values calculated from many-body perturbation theory (MBPT) \cite{yu04,martensson84} agree with the value from RCC within 3\%.
The extracted quadrupole moments are shown in Table \ref{QMtable}. The deviation of our value for $^{43}$Ca from the literature values is attributed to the low statistics of our data for this isotope combined with a poorly resolved hyperfine splitting in the excited state. Note however that our ratio of the $^{45}$Ca to $^{43}$Ca quadrupole moment is consistent with the value measured in the atomic system.

\begin{figure}[t]
\scalebox{0.7}{\input{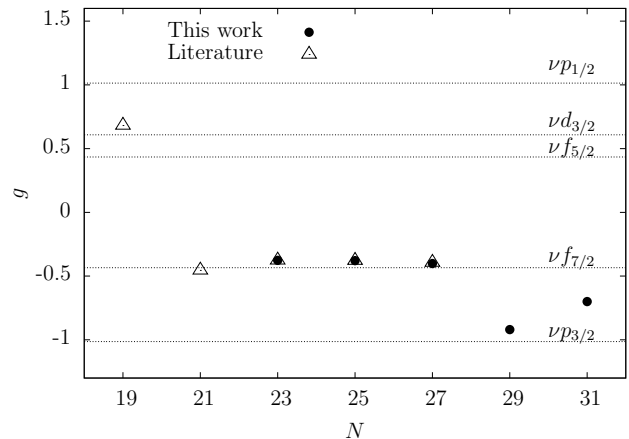}}
\vspace{-0.8cm}
\caption{Measured $g$-factors compared with literature values and 
effective single-particle values (lines) using $g_{s}^{\nu}=-3.041$
and $g_{l}^{\nu}=0.0$ ($g_{\text{eff}}^{\nu}=0.8g_{\text{free}}^{\nu}$).
The magnetic moment of $^{39}$Ca was taken from
Ref.~\cite{minamisono76}. The experimental error bars are smaller than the symbols. \label{gmoments} }
\end{figure}

Since the $g$-factors are sensitive to the valence-particle
configuration, it is illustrative to study their evolution along the Ca isotopic chain. The horizontal lines in Fig.~\ref{gmoments} show
the effective single-particle values ($g_{\text{eff}}^{\nu}=0.8g_{\text{free}}^{\nu}$) for the different shell-model orbits. The isotope $^{39}$Ca ($N=19$) has a $g$-factor close to the
$d_{3/2}$ effective single-particle value, confirming the hole nature
of this isotope. Once the $d_{3/2}$ orbit is filled, the fairly constant
$g$-factor values from $N=21$ up to $N=27$ are in agreement with that
of an odd neutron in the $f_{7/2}$ orbital.

As expected, the measured $g$-factor of $^{49}$Ca is close to the
effective single-particle value of the $p_{3/2}$ orbit, and a similar value would be expected for $^{51}$Ca. However, a deviation from this value is observed,
indicating an appreciable contribution from the mixing with configurations due to neutron excitations across $N=32$, which seems to contradict the
closed-shell nature of $N=32$. The isotope $^{51}$Ca is an exceptional case
for testing different shell-model interactions as excitations across $N=32$ can be of $M1$-type (from $p_{3/2}$ into $p_{1/2}$) and therefore even a one percent mixing of those configurations in the wave function is sufficient to induce a $\sim 20$ \% change of the $g$-factor \cite{stoyle54}.

\begin{figure}[t]
\scalebox{0.68}{\input{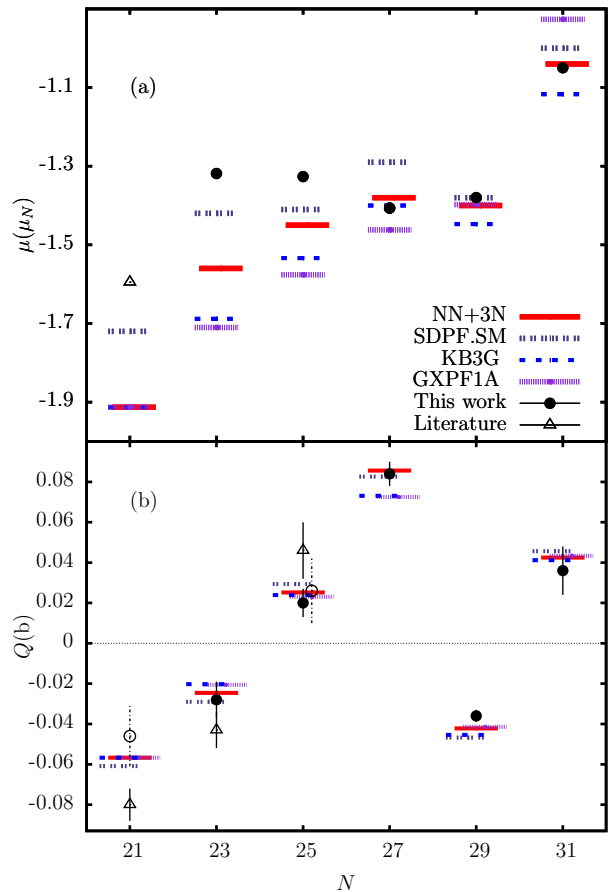}}
\vspace{-0.62cm}
\caption{Measured magnetic and quadrupole moments of Ca isotopes. Results are compared with
theoretical predictions from phenomenological interactions (KB3G,
GXPF1A, SDPF.SM) and calculations including three-nucleon forces (NN+3N). Experimental literature values (empty triangles) are given in Table~\ref{QMtable}. The open circles show the values calculated from the ratios $B(^{41}\text{Ca})/B(^{43}\text{Ca})=1.63(1)$, $B(^{45}\text{Ca})/B(^{43}\text{Ca})=-0.94(27)$  \cite{arnold83} and relative to our value of $Q(^{43}\text{Ca})$.
\label{mumoments}}
\end{figure}

The measured and calculated magnetic moments of the Ca isotopes
are shown in the upper panel of Fig.~\ref{mumoments}.  A $^{40}$Ca
core is assumed in the calculations with the GXPF1A and KB3G phenomenological interactions, as well as for the calculations with the microscopic NN+3N interaction. To investigate the effect of breaking the $^{40}$Ca core we also compare to a large-scale shell model calculation using the phenomenological interaction SDPF.SM starting from a virtual $^{28}$Si core.  For the KB3G and
GXPF1A interactions, neutrons were allowed to occupy
the $pf$ shell, while an extended valence space including the
$0g_{9/2}$ orbital ($pfg_{9/2}$ space) was used for the NN+3N
calculations. Excitations of neutrons and protons from the upper $sd$-shell into the $pf$-shell are allowed with the SDPF.SM interaction. Bare spin and orbital $g$-factors were used in all theories to
calculate the magnetic moments.  

The disagreement between the shell-model calculations starting from a $^{40}$Ca core and the
experimental magnetic moments of $^{41,43,45}$Ca suggests that nucleon excitations across the $sd$-shell are important in the vicinity of $N=20$. Indeed, large-scale shell model calculations using the SDPF.SM interaction are closer to the experimental values. These calculations include up to 6p-6h for $^{41}$Ca, 4p-4p for $^{43}$Ca, and 2p-2p for the other isotopes. A similar conclusion on the importance of cross-shell correlation across $N=20$ was obtained from experimental $g(2^+)$-factors and $B(E2)$ values of $^{42,44}$Ca \cite{schielke03,taylor05} as well as the calcium isotope shifts \cite{caurier01}. 
For the heavier Ca isotopes, all theoretical calculations describe the
experimental value rather well (see Fig.~\ref{mumoments}), indicating that from
$N=27$ and beyond, the assumption of a rigid $^{40}$Ca core works
well. Especially, the calculations with the NN+3N interaction give a very good agreement for $^{47,49,51}$Ca. Considering that from the measured $g$-factor a mixed ground state wave function is expected (Fig. \ref{gmoments}), the excellent agreement for the microscopic calculations, which are not fitted to this mass region, is remarkable. The fact that the calculated values for the phenomenological KB3G and GXPF1A lay on opposite side of the experimental value is due to the different contributions of $(p_{1/2})^2(p_{3/2})^1$ and $(p_{1/2})^1(p_{3/2})^2$ configurations. Certainly, the magnetic moment is highly sensitive to matrix elements involving the $p_{3/2}$-$p_{1/2}$ spin-orbit partners.  The ratio
of $(p_{1/2})^1(p_{3/2})^2$ to $(p_{3/2})^3$ configurations in $^{51}$Ca is a measure for these cross-shell excitations across $N=32$: it is almost twice larger with NN+3N and GXPF1A
($3.5 \%$ and $4.0 \%$, respectively) than in KB3G (2.0$\%$). Larger
cross-shell excitations reduce the absolute value of the magnetic
moment. On the other hand, due to stronger pairing, the NN+3N and KB3G interactions have a two times larger ratio of $(p_{1/2})^2(p_{3/2})^1$ over
$(p_{1/2})^1(p_{3/2})^2$ configurations than GXPF1A, 1.6 and 1.9 compared to
0.9. These cross-shell excitations increase the absolute value of the
magnetic moment.

In the lower panel of Fig.~\ref{mumoments}, the experimental
and calculated quadrupole moments are shown. The theoretical results assume neutron and proton effective charges, $e_n=0.5e$ and $e_{p}=1.5e$, respectively (protons in the valence space are allowed for the SDPF.SM interaction only). All interactions exhibit in general a good
description of the experimental quadrupole moments. The only deviation
exists for $N=27$, where KB3G and GXPF1A slightly disagree with the
experimental value, while NN+3N and SDPF.SM agrees nicely. 
The agreement between calculated and experimental values also confirms the values of effective charges used around $N=20$ \cite{caurier01,caurier07}, and more recently around $N=28$ \cite{riley14}.  Earlier studies of $N \sim Z$ isotopes, where the $f_{7/2}$ orbital is dominant, suggested values of $e_{n}=0.8e$ and $e_{p}=1.15e$ \cite{rietz04}, opening a discussion on the possible orbital dependence of the effective charges in the $pf$ shell \cite{valiente09}.

In summary, bunched-beam collinear laser spectroscopy was used to
measure the hfs spectra in the Ca II resonance transition from $^{43-51}$Ca. Our results allowed a
direct g.s.~spin determination for $^{51}$Ca. The quadrupole moments
of $^{47,49,51}$Ca, and magnetic moments of $^{49,51}$Ca were measured
for the first time. We compared these results with new shell-model calculations using a microscopic interaction derived from chiral effective field theory, including NN+3N forces, and fitted only to isotopes up to mass $A=4$. Comparison was also made with existing and new calculations using phenomenological interactions. Large discrepancies among the measured magnetic moments and the calculated values using a $^{40}$Ca core were observed around $N=20$. Large-scale shell model calculations in the $sd$-$pf$ valence space are required to reproduce better the observed magnetic moments. Further developments of microscopic interactions in the complete $sd$-$pf$ valence space for both protons and neutrons are needed to provide a consistent description for all observables of Ca isotopes from below $N=20$ up to above $N=32$. For $N \geqslant 27$, the calculations with NN+3N forces derived from chiral effective field theory provide excellent agreement for both the magnetic and quadrupole moments. 
Through the gradual filling of the $f_{7/2}$ and $p_{3/2}$ orbits, our results provide a comprehensive study of the basic ingredients employed in shell-model calculations over a wide range of neutrons. At the level of precision of our experimental results, the gs quadrupole moments do not reveal any orbital dependence of the effective charges.
The larger difference in calculated magnetic moments compared to electric quadrupole moments highlights the need of improved theoretical calculations, e.g., including two-body currents from chiral effective field theory, that compared to the present measurements may provide new insights on the magnetic operator and effective $g$-factors.
\begin{acknowledgments}

This work was supported by the IAP-project P7/12, the FWO-Vlaanderen, GOA grant 15/010 from KU Leuven, the NSF grant PHY-1068217, the BMBF Contracts Nos.~05P12RDCIC
and~06DA70471, the Max-Planck Society, the EU FP7 via ENSAR
No.~262010, the DFG through Grant SFB~634, the ERC Grant No.~307986
STRONGINT, the Helmholtz Alliance HA216/EMMI, and MINECO(SPAIN)(FPA2011-29854). Computations were
performed on JUROPA at the J\"ulich Supercomputing Center.  We would
like to thank the ISOLDE technical group for their support and
assistance.
\end{acknowledgments}

\bibliography{bibliography}

\begin{thebibliography}{53}%
\makeatletter
\providecommand \@ifxundefined [1]{%
 \@ifx{#1\undefined}
}%
\providecommand \@ifnum [1]{%
 \ifnum #1\expandafter \@firstoftwo
 \else \expandafter \@secondoftwo
 \fi
}%
\providecommand \@ifx [1]{%
 \ifx #1\expandafter \@firstoftwo
 \else \expandafter \@secondoftwo
 \fi
}%
\providecommand \natexlab [1]{#1}%
\providecommand \enquote  [1]{``#1''}%
\providecommand \bibnamefont  [1]{#1}%
\providecommand \bibfnamefont [1]{#1}%
\providecommand \citenamefont [1]{#1}%
\providecommand \href@noop [0]{\@secondoftwo}%
\providecommand \href [0]{\begingroup \@sanitize@url \@href}%
\providecommand \@href[1]{\@@startlink{#1}\@@href}%
\providecommand \@@href[1]{\endgroup#1\@@endlink}%
\providecommand \@sanitize@url [0]{\catcode `\\12\catcode `\$12\catcode
  `\&12\catcode `\#12\catcode `\^12\catcode `\_12\catcode `\%12\relax}%
\providecommand \@@startlink[1]{}%
\providecommand \@@endlink[0]{}%
\providecommand \url  [0]{\begingroup\@sanitize@url \@url }%
\providecommand \@url [1]{\endgroup\@href {#1}{\urlprefix }}%
\providecommand \urlprefix  [0]{URL }%
\providecommand \Eprint [0]{\href }%
\providecommand \doibase [0]{http://dx.doi.org/}%
\providecommand \selectlanguage [0]{\@gobble}%
\providecommand \bibinfo  [0]{\@secondoftwo}%
\providecommand \bibfield  [0]{\@secondoftwo}%
\providecommand \translation [1]{[#1]}%
\providecommand \BibitemOpen [0]{}%
\providecommand \bibitemStop [0]{}%
\providecommand \bibitemNoStop [0]{.\EOS\space}%
\providecommand \EOS [0]{\spacefactor3000\relax}%
\providecommand \BibitemShut  [1]{\csname bibitem#1\endcsname}%
\let\auto@bib@innerbib\@empty
\bibitem [{\citenamefont {Brown}(2001)}]{brown01}%
  \BibitemOpen
  \bibfield  {author} {\bibinfo {author} {\bibfnamefont {B.~A.}\ \bibnamefont
  {Brown}},\ }\href@noop {} {\bibfield  {journal} {\bibinfo  {journal} {Prog.
  Part. Nucl. Phys.}\ }\textbf {\bibinfo {volume} {47}},\ \bibinfo {pages}
  {517} (\bibinfo {year} {2001})}\BibitemShut {NoStop}%
\bibitem [{\citenamefont {Caurier}\ \emph {et~al.}(2005)\citenamefont {Caurier}
  \emph {et~al.}}]{caurier05}%
  \BibitemOpen
  \bibfield  {author} {\bibinfo {author} {\bibfnamefont {E.}~\bibnamefont
  {Caurier}} \emph {et~al.},\ }\href@noop {} {\bibfield  {journal} {\bibinfo
  {journal} {Rev. Mod. Phys.}\ }\textbf {\bibinfo {volume} {77}},\ \bibinfo
  {pages} {427} (\bibinfo {year} {2005})}\BibitemShut {NoStop}%
\bibitem [{\citenamefont {Valiente-Dobon}\ \emph {et~al.}(2009)\citenamefont
  {Valiente-Dobon} \emph {et~al.}}]{valiente09}%
  \BibitemOpen
  \bibfield  {author} {\bibinfo {author} {\bibfnamefont {J.~J.}\ \bibnamefont
  {Valiente-Dobon}} \emph {et~al.},\ }\href@noop {} {\bibfield  {journal}
  {\bibinfo  {journal} {Phys. Rev. Lett.}\ }\textbf {\bibinfo {volume} {102}},\
  \bibinfo {pages} {242502} (\bibinfo {year} {2009})}\BibitemShut {NoStop}%
\bibitem [{\citenamefont {Stetcu}\ and\ \citenamefont
  {Rotureau}(2013)}]{stetcu13}%
  \BibitemOpen
  \bibfield  {author} {\bibinfo {author} {\bibfnamefont {I.}~\bibnamefont
  {Stetcu}}\ and\ \bibinfo {author} {\bibfnamefont {J.}~\bibnamefont
  {Rotureau}},\ }\href@noop {} {\bibfield  {journal} {\bibinfo  {journal}
  {Prog. Part. Nucl. Phys.}\ }\textbf {\bibinfo {volume} {69}},\ \bibinfo
  {pages} {18} (\bibinfo {year} {2013})}\BibitemShut {NoStop}%
\bibitem [{\citenamefont {Pastore}\ \emph {et~al.}(2013)\citenamefont {Pastore}
  \emph {et~al.}}]{pastore13}%
  \BibitemOpen
  \bibfield  {author} {\bibinfo {author} {\bibfnamefont {S.}~\bibnamefont
  {Pastore}} \emph {et~al.},\ }\href@noop {} {\bibfield  {journal} {\bibinfo
  {journal} {Phys. Rev. C}\ }\textbf {\bibinfo {volume} {87}},\ \bibinfo
  {pages} {035503} (\bibinfo {year} {2013})}\BibitemShut {NoStop}%
\bibitem [{\citenamefont {Talmi}(1962)}]{talmi62}%
  \BibitemOpen
  \bibfield  {author} {\bibinfo {author} {\bibfnamefont {I.}~\bibnamefont
  {Talmi}},\ }\href@noop {} {\bibfield  {journal} {\bibinfo  {journal} {Rev.
  Mod. Phys.}\ }\textbf {\bibinfo {volume} {34}},\ \bibinfo {pages} {704}
  (\bibinfo {year} {1962})}\BibitemShut {NoStop}%
\bibitem [{\citenamefont {King}(1984)}]{kingbook}%
  \BibitemOpen
  \bibfield  {author} {\bibinfo {author} {\bibfnamefont {W.~H.}\ \bibnamefont
  {King}},\ }\href@noop {} {\emph {\bibinfo {title} {Isotope Shifts in Atomic
  Spectra}}},\ \bibinfo {edition} {1st}\ ed.\ (\bibinfo  {publisher} {Plenum
  Press. New York and London},\ \bibinfo {year} {1984})\BibitemShut {NoStop}%
\bibitem [{\citenamefont {Wienholtz}\ \emph {et~al.}(2013)\citenamefont
  {Wienholtz} \emph {et~al.}}]{wienholtz13}%
  \BibitemOpen
  \bibfield  {author} {\bibinfo {author} {\bibfnamefont {F.}~\bibnamefont
  {Wienholtz}} \emph {et~al.},\ }\href@noop {} {\bibfield  {journal} {\bibinfo
  {journal} {Nature}\ }\textbf {\bibinfo {volume} {498}},\ \bibinfo {pages}
  {7454} (\bibinfo {year} {2013})}\BibitemShut {NoStop}%
\bibitem [{\citenamefont {Steppenbeck}\ \emph {et~al.}(2013)\citenamefont
  {Steppenbeck} \emph {et~al.}}]{nature54Ca}%
  \BibitemOpen
  \bibfield  {author} {\bibinfo {author} {\bibfnamefont {D.}~\bibnamefont
  {Steppenbeck}} \emph {et~al.},\ }\href@noop {} {\bibfield  {journal}
  {\bibinfo  {journal} {Nature}\ }\textbf {\bibinfo {volume} {502}},\ \bibinfo
  {pages} {207} (\bibinfo {year} {2013})}\BibitemShut {NoStop}%
\bibitem [{\citenamefont {Holt}\ \emph {et~al.}(2012)\citenamefont {Holt} \emph
  {et~al.}}]{holt12}%
  \BibitemOpen
  \bibfield  {author} {\bibinfo {author} {\bibfnamefont {J.~D.}\ \bibnamefont
  {Holt}} \emph {et~al.},\ }\href@noop {} {\bibfield  {journal} {\bibinfo
  {journal} {J. Phys. G}\ }\textbf {\bibinfo {volume} {39}},\ \bibinfo {pages}
  {085111} (\bibinfo {year} {2012})}\BibitemShut {NoStop}%
\bibitem [{\citenamefont {Hagen}\ \emph {et~al.}(2012)\citenamefont {Hagen}
  \emph {et~al.}}]{hagen12}%
  \BibitemOpen
  \bibfield  {author} {\bibinfo {author} {\bibfnamefont {G.}~\bibnamefont
  {Hagen}} \emph {et~al.},\ }\href@noop {} {\bibfield  {journal} {\bibinfo
  {journal} {Phys. Rev. Lett.}\ }\textbf {\bibinfo {volume} {109}},\ \bibinfo
  {pages} {032502} (\bibinfo {year} {2012})}\BibitemShut {NoStop}%
\bibitem [{\citenamefont {Roth}\ \emph {et~al.}(2012)\citenamefont {Roth} \emph
  {et~al.}}]{roth12}%
  \BibitemOpen
  \bibfield  {author} {\bibinfo {author} {\bibfnamefont {R.}~\bibnamefont
  {Roth}} \emph {et~al.},\ }\href@noop {} {\bibfield  {journal} {\bibinfo
  {journal} {Phys. Rev. Lett.}\ }\textbf {\bibinfo {volume} {109}},\ \bibinfo
  {pages} {052501} (\bibinfo {year} {2012})}\BibitemShut {NoStop}%
\bibitem [{\citenamefont {Som{\`a}}\ \emph {et~al.}(2014)\citenamefont
  {Som{\`a}}, \citenamefont {Cipollone}, \citenamefont {Barbieri},
  \citenamefont {Navr{\'a}til},\ and\ \citenamefont {Duguet}}]{soma13}%
  \BibitemOpen
  \bibfield  {author} {\bibinfo {author} {\bibfnamefont {V.}~\bibnamefont
  {Som{\`a}}}, \bibinfo {author} {\bibfnamefont {A.}~\bibnamefont {Cipollone}},
  \bibinfo {author} {\bibfnamefont {C.}~\bibnamefont {Barbieri}}, \bibinfo
  {author} {\bibfnamefont {P.}~\bibnamefont {Navr{\'a}til}}, \ and\ \bibinfo
  {author} {\bibfnamefont {T.}~\bibnamefont {Duguet}},\ }\href@noop {}
  {\bibfield  {journal} {\bibinfo  {journal} {Phys. Rev. C}\ }\textbf {\bibinfo
  {volume} {89}},\ \bibinfo {pages} {061301} (\bibinfo {year}
  {2014})}\BibitemShut {NoStop}%
\bibitem [{\citenamefont {Holt}\ \emph {et~al.}(2014)\citenamefont {Holt},
  \citenamefont {Men\'endez}, \citenamefont {Simonis},\ and\ \citenamefont
  {Schwenk}}]{holt14}%
  \BibitemOpen
  \bibfield  {author} {\bibinfo {author} {\bibfnamefont {J.~D.}\ \bibnamefont
  {Holt}}, \bibinfo {author} {\bibfnamefont {J.}~\bibnamefont {Men\'endez}},
  \bibinfo {author} {\bibfnamefont {J.}~\bibnamefont {Simonis}}, \ and\
  \bibinfo {author} {\bibfnamefont {A.}~\bibnamefont {Schwenk}},\ }\href@noop
  {} {\bibfield  {journal} {\bibinfo  {journal} {Phys. Rev. C}\ }\textbf
  {\bibinfo {volume} {90}},\ \bibinfo {pages} {024312} (\bibinfo {year}
  {2014})}\BibitemShut {NoStop}%
\bibitem [{\citenamefont {Hammer}\ \emph {et~al.}(2013)\citenamefont {Hammer},
  \citenamefont {Nogga},\ and\ \citenamefont {Schwenk}}]{hammer13}%
  \BibitemOpen
  \bibfield  {author} {\bibinfo {author} {\bibfnamefont {H.-W.}\ \bibnamefont
  {Hammer}}, \bibinfo {author} {\bibfnamefont {A.}~\bibnamefont {Nogga}}, \
  and\ \bibinfo {author} {\bibfnamefont {A.}~\bibnamefont {Schwenk}},\
  }\href@noop {} {\bibfield  {journal} {\bibinfo  {journal} {Rev. Mod. Phys.}\
  }\textbf {\bibinfo {volume} {85}},\ \bibinfo {pages} {197} (\bibinfo {year}
  {2013})}\BibitemShut {NoStop}%
\bibitem [{\citenamefont {Poves}\ \emph {et~al.}(2001)\citenamefont {Poves}
  \emph {et~al.}}]{kb3}%
  \BibitemOpen
  \bibfield  {author} {\bibinfo {author} {\bibfnamefont {A.}~\bibnamefont
  {Poves}} \emph {et~al.},\ }\href@noop {} {\bibfield  {journal} {\bibinfo
  {journal} {Nucl. Phys. A}\ }\textbf {\bibinfo {volume} {694}},\ \bibinfo
  {pages} {157} (\bibinfo {year} {2001})}\BibitemShut {NoStop}%
\bibitem [{\citenamefont {Honma}\ \emph {et~al.}(2004)\citenamefont {Honma},
  \citenamefont {Otsuka}, \citenamefont {Brown},\ and\ \citenamefont
  {Mizusaki}}]{gxpf}%
  \BibitemOpen
  \bibfield  {author} {\bibinfo {author} {\bibfnamefont {M.}~\bibnamefont
  {Honma}}, \bibinfo {author} {\bibfnamefont {T.}~\bibnamefont {Otsuka}},
  \bibinfo {author} {\bibfnamefont {B.~A.}\ \bibnamefont {Brown}}, \ and\
  \bibinfo {author} {\bibfnamefont {T.}~\bibnamefont {Mizusaki}},\ }\href@noop
  {} {\bibfield  {journal} {\bibinfo  {journal} {Phys. Rev. C}\ }\textbf
  {\bibinfo {volume} {69}},\ \bibinfo {pages} {034335} (\bibinfo {year}
  {2004})}\BibitemShut {NoStop}%
\bibitem [{\citenamefont {Caurier}\ \emph {et~al.}(2007)\citenamefont {Caurier}
  \emph {et~al.}}]{caurier07}%
  \BibitemOpen
  \bibfield  {author} {\bibinfo {author} {\bibfnamefont {E.}~\bibnamefont
  {Caurier}} \emph {et~al.},\ }\href@noop {} {\bibfield  {journal} {\bibinfo
  {journal} {Phys. Rev. C}\ }\textbf {\bibinfo {volume} {75}},\ \bibinfo
  {pages} {054317} (\bibinfo {year} {2007})}\BibitemShut {NoStop}%
\bibitem [{\citenamefont {Helbert}\ \emph {et~al.}(2013)\citenamefont {Helbert}
  \emph {et~al.}}]{helbert13}%
  \BibitemOpen
  \bibfield  {author} {\bibinfo {author} {\bibfnamefont {K.}~\bibnamefont
  {Helbert}} \emph {et~al.},\ }\href@noop {} {\bibfield  {journal} {\bibinfo
  {journal} {Astrophys. J.}\ }\textbf {\bibinfo {volume} {773}},\ \bibinfo
  {pages} {11} (\bibinfo {year} {2013})}\BibitemShut {NoStop}%
\bibitem [{\citenamefont {Gallant}\ \emph {et~al.}(2012)\citenamefont {Gallant}
  \emph {et~al.}}]{gallant}%
  \BibitemOpen
  \bibfield  {author} {\bibinfo {author} {\bibfnamefont {A.~T.}\ \bibnamefont
  {Gallant}} \emph {et~al.},\ }\href@noop {} {\bibfield  {journal} {\bibinfo
  {journal} {Phys. Rev. Lett.}\ }\textbf {\bibinfo {volume} {109}},\ \bibinfo
  {pages} {032506} (\bibinfo {year} {2012})}\BibitemShut {NoStop}%
\bibitem [{\citenamefont {Isacker}\ and\ \citenamefont
  {Talmi}(2011)}]{isacker11}%
  \BibitemOpen
  \bibfield  {author} {\bibinfo {author} {\bibfnamefont {P.}~\bibnamefont
  {Isacker}}\ and\ \bibinfo {author} {\bibfnamefont {I.}~\bibnamefont
  {Talmi}},\ }\href@noop {} {\bibfield  {journal} {\bibinfo  {journal} {J.
  Phys. Conf. Ser.}\ }\textbf {\bibinfo {volume} {267}},\ \bibinfo {pages}
  {012029} (\bibinfo {year} {2011})}\BibitemShut {NoStop}%
\bibitem [{\citenamefont {Townes}\ \emph {et~al.}(1949)\citenamefont {Townes},
  \citenamefont {Foley},\ and\ \citenamefont {Low}}]{townes49}%
  \BibitemOpen
  \bibfield  {author} {\bibinfo {author} {\bibfnamefont {C.~H.}\ \bibnamefont
  {Townes}}, \bibinfo {author} {\bibfnamefont {H.~M.}\ \bibnamefont {Foley}}, \
  and\ \bibinfo {author} {\bibfnamefont {W.}~\bibnamefont {Low}},\ }\href@noop
  {} {\bibfield  {journal} {\bibinfo  {journal} {Phys. Rev.}\ }\textbf
  {\bibinfo {volume} {76}},\ \bibinfo {pages} {1415} (\bibinfo {year}
  {1949})}\BibitemShut {NoStop}%
\bibitem [{\citenamefont {Blin-Stoyle}(1956)}]{stoyle56}%
  \BibitemOpen
  \bibfield  {author} {\bibinfo {author} {\bibfnamefont {R.~J.}\ \bibnamefont
  {Blin-Stoyle}},\ }\href@noop {} {\bibfield  {journal} {\bibinfo  {journal}
  {Rev. Mod. Phys.}\ }\textbf {\bibinfo {volume} {28}},\ \bibinfo {pages} {75}
  (\bibinfo {year} {1956})}\BibitemShut {NoStop}%
\bibitem [{\citenamefont {Neyens}(2003)}]{neyens03}%
  \BibitemOpen
  \bibfield  {author} {\bibinfo {author} {\bibfnamefont {G.}~\bibnamefont
  {Neyens}},\ }\href@noop {} {\bibfield  {journal} {\bibinfo  {journal} {Rep.
  Prog. Phys.}\ }\textbf {\bibinfo {volume} {66}},\ \bibinfo {pages} {633}
  (\bibinfo {year} {2003})}\BibitemShut {NoStop}%
\bibitem [{Note1()}]{Note1}%
  \BibitemOpen
  \bibinfo {note} {A previous value for the magnetic moment of $^{49}$Ca was
  suggested from a partial measurement of its hfs and assuming a value for its
  unknown isotope shift \cite {vermeeren93}.}\BibitemShut {Stop}%
\bibitem [{\citenamefont {Fedosseev}\ \emph {et~al.}(2012)\citenamefont
  {Fedosseev} \emph {et~al.}}]{fedosseev}%
  \BibitemOpen
  \bibfield  {author} {\bibinfo {author} {\bibfnamefont {V.~N.}\ \bibnamefont
  {Fedosseev}} \emph {et~al.},\ }\href@noop {} {\bibfield  {journal} {\bibinfo
  {journal} {Rev. Sci. Inst.}\ }\textbf {\bibinfo {volume} {83}},\ \bibinfo
  {pages} {02A903} (\bibinfo {year} {2012})}\BibitemShut {NoStop}%
\bibitem [{\citenamefont {Man\'e}\ \emph {et~al.}(2009)\citenamefont {Man\'e}
  \emph {et~al.}}]{mane09}%
  \BibitemOpen
  \bibfield  {author} {\bibinfo {author} {\bibfnamefont {E.}~\bibnamefont
  {Man\'e}} \emph {et~al.},\ }\href@noop {} {\bibfield  {journal} {\bibinfo
  {journal} {Eur. Phys. J. A}\ }\textbf {\bibinfo {volume} {42}},\ \bibinfo
  {pages} {503} (\bibinfo {year} {2009})}\BibitemShut {NoStop}%
\bibitem [{\citenamefont {Papuga}\ \emph {et~al.}(2013)\citenamefont {Papuga}
  \emph {et~al.}}]{papuga13}%
  \BibitemOpen
  \bibfield  {author} {\bibinfo {author} {\bibfnamefont {J.}~\bibnamefont
  {Papuga}} \emph {et~al.},\ }\href@noop {} {\bibfield  {journal} {\bibinfo
  {journal} {Phys. Rev. Lett.}\ }\textbf {\bibinfo {volume} {110}},\ \bibinfo
  {pages} {172503} (\bibinfo {year} {2013})}\BibitemShut {NoStop}%
\bibitem [{\citenamefont {Kreim}\ \emph {et~al.}(2014)\citenamefont {Kreim}
  \emph {et~al.}}]{kim13}%
  \BibitemOpen
  \bibfield  {author} {\bibinfo {author} {\bibfnamefont {K.}~\bibnamefont
  {Kreim}} \emph {et~al.},\ }\href@noop {} {\bibfield  {journal} {\bibinfo
  {journal} {Phys. Lett. B}\ }\textbf {\bibinfo {volume} {731}},\ \bibinfo
  {pages} {97} (\bibinfo {year} {2014})}\BibitemShut {NoStop}%
\bibitem [{\citenamefont {Kowalska}\ \emph {et~al.}(2008)\citenamefont
  {Kowalska} \emph {et~al.}}]{kowalska}%
  \BibitemOpen
  \bibfield  {author} {\bibinfo {author} {\bibfnamefont {M.}~\bibnamefont
  {Kowalska}} \emph {et~al.},\ }\href@noop {} {\bibfield  {journal} {\bibinfo
  {journal} {Phys. Rev.}\ }\textbf {\bibinfo {volume} {C 77}},\ \bibinfo
  {pages} {034307} (\bibinfo {year} {2008})}\BibitemShut {NoStop}%
\bibitem [{\citenamefont {Bergmann}\ \emph {et~al.}(1980)\citenamefont
  {Bergmann} \emph {et~al.}}]{bergman80}%
  \BibitemOpen
  \bibfield  {author} {\bibinfo {author} {\bibfnamefont {E.}~\bibnamefont
  {Bergmann}} \emph {et~al.},\ }\href@noop {} {\bibfield  {journal} {\bibinfo
  {journal} {Z. Phys. A}\ }\textbf {\bibinfo {volume} {294}},\ \bibinfo {pages}
  {319} (\bibinfo {year} {1980})}\BibitemShut {NoStop}%
\bibitem [{\citenamefont {Arnold}\ \emph {et~al.}(1983)\citenamefont {Arnold}
  \emph {et~al.}}]{arnold83}%
  \BibitemOpen
  \bibfield  {author} {\bibinfo {author} {\bibfnamefont {M.}~\bibnamefont
  {Arnold}} \emph {et~al.},\ }\href@noop {} {\bibfield  {journal} {\bibinfo
  {journal} {Z. Phys, A}\ }\textbf {\bibinfo {volume} {314}},\ \bibinfo {pages}
  {303} (\bibinfo {year} {1983})}\BibitemShut {NoStop}%
\bibitem [{\citenamefont {Arbes}\ \emph {et~al.}(1994)\citenamefont {Arbes}
  \emph {et~al.}}]{arbes}%
  \BibitemOpen
  \bibfield  {author} {\bibinfo {author} {\bibfnamefont {F.}~\bibnamefont
  {Arbes}} \emph {et~al.},\ }\href@noop {} {\bibfield  {journal} {\bibinfo
  {journal} {Z. Phys. D.}\ }\textbf {\bibinfo {volume} {31}},\ \bibinfo {pages}
  {27} (\bibinfo {year} {1994})}\BibitemShut {NoStop}%
\bibitem [{\citenamefont {Silverans}\ \emph {et~al.}(1991)\citenamefont
  {Silverans} \emph {et~al.}}]{silveransZD}%
  \BibitemOpen
  \bibfield  {author} {\bibinfo {author} {\bibfnamefont {R.~E.}\ \bibnamefont
  {Silverans}} \emph {et~al.},\ }\href@noop {} {\bibfield  {journal} {\bibinfo
  {journal} {Z. Phys. D.}\ }\textbf {\bibinfo {volume} {18}},\ \bibinfo {pages}
  {351} (\bibinfo {year} {1991})}\BibitemShut {NoStop}%
\bibitem [{\citenamefont {N{\"o}rtershauser}\ \emph {et~al.}(1998)\citenamefont
  {N{\"o}rtershauser} \emph {et~al.}}]{norte98}%
  \BibitemOpen
  \bibfield  {author} {\bibinfo {author} {\bibfnamefont {W.}~\bibnamefont
  {N{\"o}rtershauser}} \emph {et~al.},\ }\href@noop {} {\bibfield  {journal}
  {\bibinfo  {journal} {Eur. Phys. J. D}\ }\textbf {\bibinfo {volume} {2}},\
  \bibinfo {pages} {33} (\bibinfo {year} {1998})}\BibitemShut {NoStop}%
\bibitem [{\citenamefont {Perrot}\ \emph {et~al.}(2006)\citenamefont {Perrot}
  \emph {et~al.}}]{perrot06}%
  \BibitemOpen
  \bibfield  {author} {\bibinfo {author} {\bibfnamefont {F.}~\bibnamefont
  {Perrot}} \emph {et~al.},\ }\href@noop {} {\bibfield  {journal} {\bibinfo
  {journal} {Phys. Rev. C}\ }\textbf {\bibinfo {volume} {74}},\ \bibinfo
  {pages} {014313} (\bibinfo {year} {2006})}\BibitemShut {NoStop}%
\bibitem [{\citenamefont {Fornal}\ \emph {et~al.}(2008)\citenamefont {Fornal}
  \emph {et~al.}}]{fornal08}%
  \BibitemOpen
  \bibfield  {author} {\bibinfo {author} {\bibfnamefont {B.}~\bibnamefont
  {Fornal}} \emph {et~al.},\ }\href@noop {} {\bibfield  {journal} {\bibinfo
  {journal} {Phys. Rev. C}\ }\textbf {\bibinfo {volume} {77}},\ \bibinfo
  {pages} {014304} (\bibinfo {year} {2008})}\BibitemShut {NoStop}%
\bibitem [{\citenamefont {Olschewski}(1972)}]{olschewski72}%
  \BibitemOpen
  \bibfield  {author} {\bibinfo {author} {\bibfnamefont {L.}~\bibnamefont
  {Olschewski}},\ }\href@noop {} {\bibfield  {journal} {\bibinfo  {journal} {Z.
  Physik}\ }\textbf {\bibinfo {volume} {249}},\ \bibinfo {pages} {205}
  (\bibinfo {year} {1972})}\BibitemShut {NoStop}%
\bibitem [{\citenamefont {Brun}\ \emph {et~al.}(1962)\citenamefont {Brun} \emph
  {et~al.}}]{brun62}%
  \BibitemOpen
  \bibfield  {author} {\bibinfo {author} {\bibfnamefont {E.}~\bibnamefont
  {Brun}} \emph {et~al.},\ }\href@noop {} {\bibfield  {journal} {\bibinfo
  {journal} {Phys. Rev. Lett.}\ }\textbf {\bibinfo {volume} {9}},\ \bibinfo
  {pages} {166} (\bibinfo {year} {1962})}\BibitemShut {NoStop}%
\bibitem [{\citenamefont {Sundholm}\ and\ \citenamefont
  {Olsen}(1993)}]{sundholm}%
  \BibitemOpen
  \bibfield  {author} {\bibinfo {author} {\bibfnamefont {D.}~\bibnamefont
  {Sundholm}}\ and\ \bibinfo {author} {\bibfnamefont {J.}~\bibnamefont
  {Olsen}},\ }\href@noop {} {\bibfield  {journal} {\bibinfo  {journal} {J.
  Chem. Phys.}\ }\textbf {\bibinfo {volume} {98}},\ \bibinfo {pages} {7152}
  (\bibinfo {year} {1993})}\BibitemShut {NoStop}%
\bibitem [{\citenamefont {Sahoo}(2009)}]{sahoo09}%
  \BibitemOpen
  \bibfield  {author} {\bibinfo {author} {\bibfnamefont {B.~K.}\ \bibnamefont
  {Sahoo}},\ }\href@noop {} {\bibfield  {journal} {\bibinfo  {journal} {Phys.
  Rev. A}\ }\textbf {\bibinfo {volume} {80}},\ \bibinfo {pages} {012515}
  (\bibinfo {year} {2009})}\BibitemShut {NoStop}%
\bibitem [{\citenamefont {Arnold}\ \emph {et~al.}(1981)\citenamefont {Arnold}
  \emph {et~al.}}]{arnold81}%
  \BibitemOpen
  \bibfield  {author} {\bibinfo {author} {\bibfnamefont {M.}~\bibnamefont
  {Arnold}} \emph {et~al.},\ }\href@noop {} {\bibfield  {journal} {\bibinfo
  {journal} {Hyperfine Interact.}\ }\textbf {\bibinfo {volume} {9}},\ \bibinfo
  {pages} {159} (\bibinfo {year} {1981})}\BibitemShut {NoStop}%
\bibitem [{\citenamefont {Andl}\ \emph {et~al.}(1982)\citenamefont {Andl} \emph
  {et~al.}}]{andl82}%
  \BibitemOpen
  \bibfield  {author} {\bibinfo {author} {\bibfnamefont {A.}~\bibnamefont
  {Andl}} \emph {et~al.},\ }\href@noop {} {\bibfield  {journal} {\bibinfo
  {journal} {Phys. Rev. C}\ }\textbf {\bibinfo {volume} {26}},\ \bibinfo
  {pages} {2194} (\bibinfo {year} {1982})}\BibitemShut {NoStop}%
\bibitem [{\citenamefont {Kai-zhi}\ \emph {et~al.}(2004)\citenamefont {Kai-zhi}
  \emph {et~al.}}]{yu04}%
  \BibitemOpen
  \bibfield  {author} {\bibinfo {author} {\bibfnamefont {Y.}~\bibnamefont
  {Kai-zhi}} \emph {et~al.},\ }\href@noop {} {\bibfield  {journal} {\bibinfo
  {journal} {Phys. Rev. A}\ }\textbf {\bibinfo {volume} {70}},\ \bibinfo
  {pages} {012506} (\bibinfo {year} {2004})}\BibitemShut {NoStop}%
\bibitem [{\citenamefont {Martensson-Pendrill}\ and\ \citenamefont
  {Salomonson}(1984)}]{martensson84}%
  \BibitemOpen
  \bibfield  {author} {\bibinfo {author} {\bibfnamefont {A.~M.}\ \bibnamefont
  {Martensson-Pendrill}}\ and\ \bibinfo {author} {\bibfnamefont
  {S.}~\bibnamefont {Salomonson}},\ }\href@noop {} {\bibfield  {journal}
  {\bibinfo  {journal} {Phys. Rev. A}\ }\textbf {\bibinfo {volume} {30}},\
  \bibinfo {pages} {712} (\bibinfo {year} {1984})}\BibitemShut {NoStop}%
\bibitem [{\citenamefont {Minamisono}\ \emph {et~al.}(1976)\citenamefont
  {Minamisono} \emph {et~al.}}]{minamisono76}%
  \BibitemOpen
  \bibfield  {author} {\bibinfo {author} {\bibfnamefont {T.}~\bibnamefont
  {Minamisono}} \emph {et~al.},\ }\href@noop {} {\bibfield  {journal} {\bibinfo
   {journal} {Phys. Lett. B}\ }\textbf {\bibinfo {volume} {61}},\ \bibinfo
  {pages} {155} (\bibinfo {year} {1976})}\BibitemShut {NoStop}%
\bibitem [{\citenamefont {Blin-Stoyle}\ and\ \citenamefont
  {Perks}(1954)}]{stoyle54}%
  \BibitemOpen
  \bibfield  {author} {\bibinfo {author} {\bibfnamefont {R.~J.}\ \bibnamefont
  {Blin-Stoyle}}\ and\ \bibinfo {author} {\bibfnamefont {M.~A.}\ \bibnamefont
  {Perks}},\ }\href@noop {} {\bibfield  {journal} {\bibinfo  {journal} {Proc.
  Phys. Soc.}\ }\textbf {\bibinfo {volume} {A67}},\ \bibinfo {pages} {885}
  (\bibinfo {year} {1954})}\BibitemShut {NoStop}%
\bibitem [{\citenamefont {Schielke}\ \emph {et~al.}(2003)\citenamefont
  {Schielke} \emph {et~al.}}]{schielke03}%
  \BibitemOpen
  \bibfield  {author} {\bibinfo {author} {\bibfnamefont {S.}~\bibnamefont
  {Schielke}} \emph {et~al.},\ }\href@noop {} {\bibfield  {journal} {\bibinfo
  {journal} {Phys. Lett. B}\ }\textbf {\bibinfo {volume} {571}},\ \bibinfo
  {pages} {29} (\bibinfo {year} {2003})}\BibitemShut {NoStop}%
\bibitem [{\citenamefont {Taylor}\ \emph {et~al.}(2005)\citenamefont {Taylor}
  \emph {et~al.}}]{taylor05}%
  \BibitemOpen
  \bibfield  {author} {\bibinfo {author} {\bibfnamefont {M.~J.}\ \bibnamefont
  {Taylor}} \emph {et~al.},\ }\href@noop {} {\bibfield  {journal} {\bibinfo
  {journal} {Phys. Lett. B}\ }\textbf {\bibinfo {volume} {605}},\ \bibinfo
  {pages} {265} (\bibinfo {year} {2005})}\BibitemShut {NoStop}%
\bibitem [{\citenamefont {Caurier}\ \emph {et~al.}(2001)\citenamefont {Caurier}
  \emph {et~al.}}]{caurier01}%
  \BibitemOpen
  \bibfield  {author} {\bibinfo {author} {\bibfnamefont {E.}~\bibnamefont
  {Caurier}} \emph {et~al.},\ }\href@noop {} {\bibfield  {journal} {\bibinfo
  {journal} {Phys. Lett. B}\ }\textbf {\bibinfo {volume} {522}},\ \bibinfo
  {pages} {3} (\bibinfo {year} {2001})}\BibitemShut {NoStop}%
\bibitem [{\citenamefont {Riley}\ \emph {et~al.}(2014)\citenamefont {Riley}
  \emph {et~al.}}]{riley14}%
  \BibitemOpen
  \bibfield  {author} {\bibinfo {author} {\bibfnamefont {L.~A.}\ \bibnamefont
  {Riley}} \emph {et~al.},\ }\href@noop {} {\bibfield  {journal} {\bibinfo
  {journal} {Phys. Rev. C}\ }\textbf {\bibinfo {volume} {90}},\ \bibinfo
  {pages} {011305} (\bibinfo {year} {2014})}\BibitemShut {NoStop}%
\bibitem [{\citenamefont {du~Rietz}\ \emph {et~al.}(2004)\citenamefont
  {du~Rietz} \emph {et~al.}}]{rietz04}%
  \BibitemOpen
  \bibfield  {author} {\bibinfo {author} {\bibfnamefont {R.}~\bibnamefont
  {du~Rietz}} \emph {et~al.},\ }\href@noop {} {\bibfield  {journal} {\bibinfo
  {journal} {Phys. Rev. Lett.}\ }\textbf {\bibinfo {volume} {93}},\ \bibinfo
  {pages} {222501} (\bibinfo {year} {2004})}\BibitemShut {NoStop}%
\bibitem [{\citenamefont {Vermeeren}\ \emph {et~al.}(1993)\citenamefont
  {Vermeeren} \emph {et~al.}}]{vermeeren93}%
  \BibitemOpen
  \bibfield  {author} {\bibinfo {author} {\bibfnamefont {L.}~\bibnamefont
  {Vermeeren}} \emph {et~al.},\ }\href@noop {} {\bibfield  {journal} {\bibinfo
  {journal} {Proc. of 6th Int. Conf. on Nucl. far from Stability}\ ,\ \bibinfo
  {pages} {193}} (\bibinfo {year} {1993})}\BibitemShut {NoStop}%
\end{thebibliography}%

\end{document}